\documentstyle[amsmath,amssymb,multicol,graphics,aps]{revtex}       %draft,overcite,

\begin{document}

\title{Angle-Resolved Spectroscopy of Electron-Electron Scattering in a 2D System}%

  \author{A.V. Yanovsky$\star$, H. Predel \dag, H. Buhmann \dag, R.N. Gurzhi $\star$,
  A.N. Kalinenko $\star$, A.I. Kopeliovich $\star$, and L.W. Molenkamp \dag}

  %\affiliation{
  \address{
  \dag Physikalisches Institut der Universit\"at W\"urzburg, D-97047
  W\"urzburg, Germany. \\ $\star$B.Verkin Institute for Low Temperature Physics
  \mbox{\&} Engineering, \\ Nat.Acad. of Sciences of Ukraine, Lenin Ave. 47,
  31064,
  Kharkov, Ukraine}

\maketitle

\begin{abstract}

Electron-beam propagation experiments have been used to determine
the energy and angle dependence of electron-electron (ee)
scattering a two-dimensional electron gas (2DEG) in a very direct
manner by a new spectroscopy method. The experimental results are
in good agreement with recent theories and provide direct evidence
for the differences between ee-scattering in a 2DEG as compared
with 3D systems. Most conspicuous is the increased importance of
small-angle scattering in a 2D system, resulting in a reduced (but
energy-dependent) broadening of the electron beam.

72.10.-d,72.20.Dp
\end{abstract}

%%%%%%%%%%%%%%%%%%%%%%%%%%
\begin{multicols}{2}
%%%%%%%%%%%%%%%%%%%%%%%%%%%

The scattering characteristics of electrons in systems with
reduced dimensions are expected to exhibit decisive differences
with respect to the situation in the bulk. Theoretically,
electron-electron (ee) scattering in two-dimensional (2D) systems
was first considered in the early seventies~\cite{chaplikhodges}.
Further numerical evaluations followed in the early
eighties~\cite{guiliani82}. It was shown that the life-time of a
non-equilibrium electron in a 2DEG is shorter by a factor of order
$\ln(\varepsilon_F/ \varepsilon)$ compared to the
three-dimensional (3D) case ($\varepsilon$ is the electron's
excess energy counted from the Fermi energy, $\varepsilon_{F}$).

A reduction of the dimensionality induces much more drastic
changes in the momentum transfer
processes~\cite{gurzhi87,gprb,gfnt97,Laikhtman,buhmann98}. Two
types of ee-collisions with nearly the same probability
characterize scattering in 2D systems~\cite{gurzhi87}: (i)
Collisions of a non-equilibrium electron with momentum $\bf p$ and
excess energy $\varepsilon$ with equilibrium electrons $\bf p_{1}$
usually result in scattering of both electron by a small angle
$\alpha \sim \varepsilon/ \varepsilon_{F}$ into states $\bf p_{2}$
and $\bf p_{3}$ leaving a hole (an empty place in Fermi
distribution) in state $\bf p_{1}$, with $\bf p + p_{1} =
p_{2}+p_{3}$. (ii) Collisions with electrons of nearly opposite
momentum, $\bf p \approx - p_{1}$: In this case, the electrons at
$\bf p_{2}$ and $\bf p_{3}\approx - p_{2}$ are scattered by a much
larger angle, on average
$\alpha\approx\sqrt{\varepsilon/\varepsilon_F}$.

\begin{figure}[H]
\begin{center}
\resizebox{6.25cm}{5cm}{\includegraphics{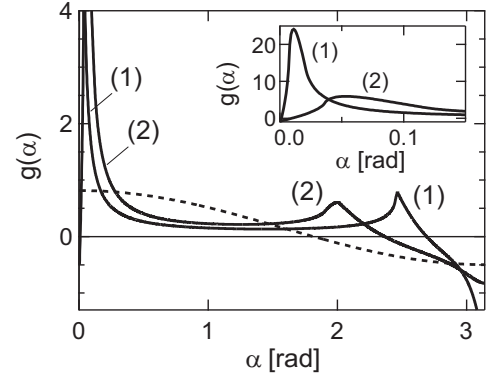}}
\end{center}
\caption{Electron-electron  scattering angular distribution
function $g(\alpha)$ in a 2D System, $T=0$.
(1)$\varepsilon=0.12\varepsilon_{F}$,
(2)$\varepsilon=0.4\varepsilon_{F}$,
 dashed: 3D case (Callaway's Ansatz).}
    \label{indicatrix}
\end{figure}

The details of the scattering behaviour can be visualized by the
angular distribution function of the scattered electrons,
$g(\alpha)$, where the angle $\alpha$ is measured with respect to
$\bf p$~\cite{buhmann98}. By definition, $|g(\alpha)|d\alpha$ is
the probability that a non-equilibrium electron, $g(\alpha)>0$ [or
hole for $g(\alpha)<0$], emerges in an interval $d\alpha$ after
scattering. This function $g(\alpha)$ is shown for two different
electron excess energies $\varepsilon$ in Fig.~\ref{indicatrix}.
For comparison, we have also plotted the most commonly used
approximation for $g(\alpha)$ for 3D systems, sometimes referred
to as the Callaway Ansatz [$g(\alpha) \propto 1 + 2 \cos(\alpha)$,
and independent of $\varepsilon$]. For 3D systems $g(\alpha)$ is
very smooth, exhibiting a broad distribution of electrons moving
in forward direction and holes moving backwards. In the 2D case
$g(\alpha)$ shows several distinct features. Most conspicuous is a
very narrow distribution of electrons moving in forward direction.
%%%%%%%%%%%%%%%%%%%%%% Fig. 1 %%%%%%%%%%%%%%%%%%%%%%%%%
The height of this peak is determined by the small-angular
processes of type (i); its width is determined by type (ii)
processes and increases with energy according to
$\delta\alpha\sim\sqrt{\varepsilon/\varepsilon_F}$. The type (ii)
scattering events also cause a secondary peak at
$\alpha\approx\pi-2\sqrt{\varepsilon/\varepsilon_F}$ and a narrow
hole dip of width $\sqrt{\varepsilon/\varepsilon_F}$ at
$\alpha=\pi$ \cite{buhmann98}. However, these effects occur in the
backscattered direction and are quite small, so that they will be
difficult to detect experimentally. For a comparison with
experiments we therefore focus on angles $\alpha<1\,\,\,$(rad),
where the small angular scattering peak should provide a clear
token of specific 2D phenomena. Another intriguing feature of $g$,
which can be seen more clearly in the inset of
Fig.~\ref{indicatrix}, is a dip in forward direction for {\it
very} small angles, with a width $\sim
0.1(\varepsilon/\varepsilon_F)^{3/2}$. This dip is caused by the
conservation laws: The electron may give away its surplus energy
to equilibrium partners only upon scattering by a finite angle.
This effect, which was discussed earlier in Refs.
\cite{gprb,cumming96},  also occurs in 3D systems. However, in 2D
the amplitude of the dip is enhanced by a factor
$\varepsilon_F/\varepsilon$.

Up to now no direct experimental evidence for these 2D effects in
ee-scattering has been demonstrated. Recently,
%%%%%%%%%%%%%%%%%%%%%% Fig. 2 %%%%%%%%%%%%%%%%%%%%%%%%%
\begin{figure}[T]
\begin{center}
\resizebox{6.25cm}{5.0cm}{\includegraphics{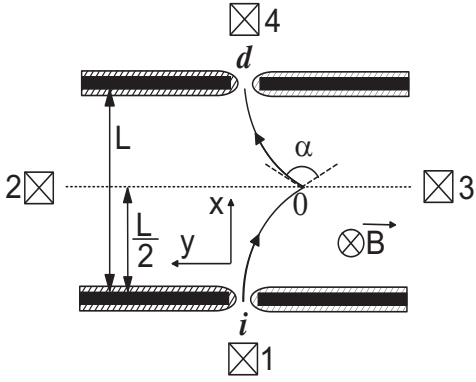}}
\end{center}
\caption{Schematic view of the sample structure showing the
Schottky gates (black areas) defining the injector ($i$) and
detector ($d$) point contacts. Also indicated is a possible
trajectory of an injected electron in a perpindicular magnetic
field $B$, where the electron is scattered in point $O$ over an
angle $\alpha$. Hatched areas and crosses represent the depleted
2DEG-regions and ohmic contacts, respectively.}
    \label{sample}
\end{figure}
we published results indicating the dominance of small angle
scattering in the propagation of an electron beam in a 2DEG
indirectly \cite{predel00}.
%Extending the experimental techniques used in that paper,
Now, we have been able to extract $g(\alpha)$ directly. This
experiment provides compelling evidence for the preponderance of
small-angle scattering in 2D systems.

In the experiment, an electron beam injected into the 2DEG via an
electrostatically defined quantum point-contact (QPC) {\it i} is
detected by a second QPC {\it d} in a certain distance
\cite{Mol90,Heiblum}, schematically shown in Fig.~\ref{sample}.
When a magnetic field is applied perpendicular to the 2DEG plane,
the injected beam is deflected and only scattered electrons can
reach the detector QPC. At low electron excess energies
$\varepsilon \ll \varepsilon_{F}$ one can neglect the energy
dependence of the cyclotron radius $r_{c}$. When the opening angle
$\Phi$ of the QPCs $i$ and $d$ \cite{Mol90} is sufficiently small
i.e., $\Phi \ll 1$, we have that for a given magnetic field $B$
the detector signal is determined only by one trajectory i.e., the
signal results solely from electrons that were scattered in point
$O$ across an angle $\alpha=2\arcsin(L/2r_{c})$ (see
Fig.~\ref{sample}). Thus, by changing the magnetic field we can
directly measure the angular distribution function of scattered
electrons $g(\alpha)$ in a wide range of angles $\alpha$.

As discussed above, $g(\alpha)$ will in general depend on the
excess energy $\varepsilon$ of the injected electrons. This is why
we apply a differential measurement technique, which is equivalent
to using mono-energetic electron beams. $\varepsilon$ is
controlled by adjusting the bias voltage $V_i$ applied between
contacts 1 and 2. The non-local voltage drop $V_d$ measured
between contacts 3 and 4 results from electrons have reached the
detector QPC and charge the 2DEG area denoted $d$. A small
$ac$-modulation $\delta V_i\ll V_i$ is added to the $dc$-bias.
Although an electron beam injected via a QPC consists of electrons
of all energies form $\varepsilon_F$ up to
$\varepsilon_F+\varepsilon$, only the contribution $\delta V_d$ of
the high energy part of the beam to the signal can be detected by
measuring %%%%%%%%%%%%%%%%%%%%%% Fig. 3 %%%%%%%%%%%%%%%%%%%%%%%%%
\begin{figure}
\begin{center}
\resizebox{6.25cm}{5cm}{\includegraphics{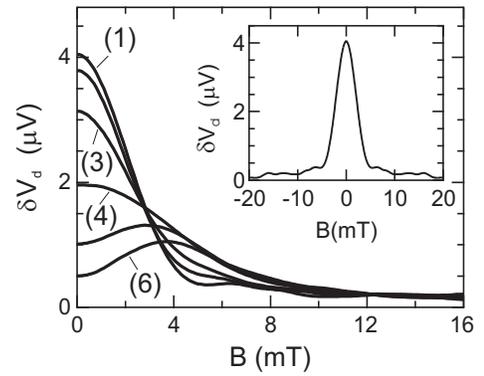}}
\end{center}
\caption{a) Behaviour of the electron beam signal at different
injector voltages: (1) $V_i = 0.8$ mV, (2) $V_i =1.2$ mV, (3) $V_i
=1.6$ mV, (4) $V_i =2.6$ mV, (5) $V_i=3.5$ mV and (6) $V_i=4.5$ mV
as a function of magnetic field $B$. Inset: curve for $V_i=0.1$
mV.}
    \label{measurement}
\end{figure}
the signal with a lock-in at the same frequency as $\delta V_i$.

For the experiments, conventional Si-modulation doped
GaAs-(Ga,Al)As heterosturctures were used, with a carrier
concentration of $n_s \simeq 2.8 \times 10^{11}$ cm$^{-2}$ and an
electron mobility of $\mu \simeq 100$ m$^2$(V s)$^{-1}$, which
implies an impurity mean free path of $l_{\rm imp} \geq 10$
$\mu$m. A pair of QPCs, about $L\approx 4$~$\mu$m apart, were
fabricated using split-gate technology. By applying a negative
voltage to the gate contacts, the conductance of the QPCs
($G_{QPC} = N 2e^2/h $) could be adjusted from several conducting
modes $N$ into the tunneling regime ($N < 1$). Throughout all
experiments injector and detector QPC were adjusted to $N = 1$ to
ensure narrow opening angles~\cite{Mol90}. The injection
dc-voltage, $V_i = V_{12}$, was varied between 0 and 5 mV. The
ac-modulation voltage was kept constant at $30\,\,\mu$V, so that
$\delta V_i \leq k_BT_0/e \ll V_i$. The sample was kept at a
lattice temperature  $T_0 \approx 200$ mK in a dilution
refrigerator.

Fig.~\ref{measurement} displays some examples of the measured
detector signal for various injection voltages as a function of
magnetic field. The inset of Fig.~\ref{measurement} shows the
measured signal at low injection energy $eV_i=0.1$\ meV, when the
ee-scattering mean free path $l_{ee}$ is much larger than $L$ and
the electrons reach the detector QPC ballistically. From this we
determine the characteristic opening angle \cite{Mol90} of
injector and detector, $\Phi\approx 12^{\circ}$. The detector
signal is at maximum at zero magnetic field. With increasing
injection energy $V_i$ ee-scattering becomes more important,
leading to (i) a decrease of the signal of non-scattered electrons
near $B=0$, (ii) a broadening of the signal with $B$ and (iii) the
appearance of a dip in the signal around $B=0$ for energies $V_i
\ge 3.5$ mV.

For further consideration we have to investigate how this
experimental behaviour relates quantitatively to the expected
2D-scattering characteristics. Therefore, we describe the problem
by a linearized Boltzmann equation in magnetic field,
%%%%%%%%%%%%%%%%%% Eq. 1 %%%%%%%%%%%%%%%%%%%%%%%%%%%%%%
  \begin{equation}\label{e-kin}
    \omega_{c} \partial f/ \partial \varphi+v_{x}\partial f/
    \partial x + v_{y}\partial f/\partial y=\hat{J}f \; ,
  \end{equation}
%%%%%%%%%%%%%%%%%%%%%%%%%%%%%%%%%%%%%%%%%%%%%%%%%%%%%%%
where $\omega_c$ is the cyclotron frequency and $\hat{J}$ is the
linearized operator of the ee-collisions, which can be written as:
%%%%%%%%%%%%%%%%%% Eq. 2 %%%%%%%%%%%%%%%%%%%%%%%%%%%%%%
  \begin{equation} \label{e-colint}
    \hat{J}f({\bf p})=-\nu f({\bf p})+\int d{\bf p'}\,\nu_{{\bf p}
    {\bf p'}}\,f({\bf p'})\, ,
  \end{equation}
%%%%%%%%%%%%%%%%%%%%%%%%%%%%%%%%%%%%%%%%%%%%%%%%%%%%%%%
with $\nu=\int d{\bf p'} \, \nu_{{\bf p'}{\bf p}}$. Integration of
the collision integral kernel $\nu_{{\bf p'p}}$ over energy yields
the angular distribution function of the scattered electrons:
%%%%%%%%%%%%%%%%%% Eq. 3 %%%%%%%%%%%%%%%%%%%%%%%%%%%%%%
  \begin{equation} \label{e-gdef}
    g(\alpha)=m\nu^{-1}\,\int d\varepsilon^{\prime} \, \nu_{\bf
    p^{\prime}p}\, ,
  \end{equation}
%%%%%%%%%%%%%%%%%%%%%%%%%%%%%%%%%%%%%%%%%%%%%%%%%%%%%%%
where $\alpha$ is the angle between ${\bf p}$ and ${\bf p'}$, and
${\bf p'}$ refers to the electrons (holes) at ${\bf p}_1$, ${\bf
p}_2$ and ${\bf p}_3$ mentioned above. If the probability for an
electron to be scattered over a distance $L$ is small (i.e.,
$l_{ee}(\varepsilon) = v \cdot \nu^{-1} \gg L$), Eq.~\ref{e-kin}
can be solved using pertubation theory on the collision integral.
When we write the electron distribution function at the exit of
the injector as $f_{0} = \delta V_i \, \delta (\varepsilon -
V_i)\, \lambda_F \delta(y)\rho_{i}(\varphi)$, and consider only
the first iteration of the collision integral, we can obtain an
expression for the current through the detector QPC. For low
injection energies $eV_i = \varepsilon \ll \varepsilon_F$, the
detector signal can be written as:
%%%%%%%%%%%%%%%%%% Eq. 4 %%%%%%%%%%%%%%%%%%%%%%%%%%%%%%
  \begin{eqnarray}\label{e-genVs}
    \delta V_{d}^{s} &\approx & C \, \nu \, \int\limits_{-\pi/2}^{\pi/2}d \varphi
    \int\limits_{\varphi_0}^{\varphi}d \varphi'' \int\limits_{-\pi/2}^{\pi/2}d\varphi' \,
    \rho_{d}(\varphi)\,\rho_{i}(\varphi_{1})\cos \varphi \times
    \nonumber \\
    & \times & g(\varphi''-\varphi',V_{i})
    \delta(\cos\varphi+\cos\varphi'-\cos
    \varphi''-\cos\varphi_1),\nonumber \\
    && \varphi_{0} =\arcsin(\sin\varphi-L /
    r_{c}), \nonumber \\
    && \varphi_1  = \arcsin
    (\sin \varphi' - \sin \varphi''+ \sin \varphi_0).
  \end{eqnarray} % \end{multline}
%%%%%%%%%%%%%%%%%%%%%%%%%%%%%%%%%%%%%%%%%%%%%%%%%%%%%%%
Here $C = 2 m L \lambda_F \delta V_i (h \, e)^{-1}$, $\lambda_F$
is the Fermi wave length and $\rho_{i}(\varphi)$ [
$\rho_{d}(\varphi)$] the angular emittance (acceptance) function
of the injector (detector) QPC \cite{Mol90}. From this equation it
is clear that $g(\alpha)$ can be obtained from the magnetic field
dependence of $\delta V_d^s$.

When $g(\alpha)$ varies only slightly on the scale of the opening
angle $\Phi$, a local approximation to the integrals in
Eq.(\ref{e-genVs}) can be made, yielding
%%%%%%%%%%%%%%%%%% Eq. 5 %%%%%%%%%%%%%%%%%%%%%%%%%%%%%%
  \begin{equation} \label{e-spectVi}
    \delta V_{d}^{s}=2 \, C \, \nu \, K(\alpha, \Phi )
    g(\alpha, V_{i}), \,\,\, \alpha = 2 \arcsin L/2r_c.
  \end{equation}
%%%%%%%%%%%%%%%%%%%%%%%%%%%%%%%%%%%%%%%%%%%%%%%%%%%%%%%
The factor $K$ is given by $K \approx r_c/L \sqrt{1-(L/r_c)^2}$
for $\Phi < \alpha < \pi - 2 \sqrt{\Phi}$ and $K \approx 1/\Phi$
for $\alpha <\Phi$. Note that for small enough beam energies
$V_i$, a local approximation of Eq.~(\ref{e-spectVi}) is invalid
for scattering angles $\alpha \lesssim \Phi$ and
Eq.~(\ref{e-spectVi}) only yields a smoothed (by the emittance and
acceptance functions) approximation of $g(\alpha)$. The local
approximation is also invalid for large scattering angles, $\pi -
\alpha < 2\sqrt{\Phi}$.

However, it is possible to extend the range of validity for this
one-collision approximation. This is because in all experiments we
have that $V_i \gg T$, implying that the probability for secondary
ee-collisions is approximately an order of magnitude lower than
that of the first one~\cite{gfnt97,buhmann98,gssc}. It turns out
that a one-collision approximation is valid as long as
$L<l_{ee}(eV_i /3)\approx 10\,l_{ee}(V_i)$, i.e. for a much wider
range of parameters than the perturbation theory. Partial
summation of the corresponding iteration series of
Eq.~(\ref{e-kin}), results in the following expression for
$\tilde{g}(\alpha,V_i)$:
%%%%%%%%%%%%%%%%%% Eq. 6 %%%%%%%%%%%%%%%%%%%%%%%%%%%%%%
   \begin{equation}\label{e-mocaG}
    \tilde{g}(\alpha, V_i )=\exp \left(
    -\frac{\Lambda}{l_{ee}}\right) g(\alpha, V_i )\, ,
   \end{equation}
%%%%%%%%%%%%%%%%%%%%%%%%%%%%%%%%%%%%%%%%%%%%%%%%%%%%%%%
which replaces $g(\alpha, V_i)$ in Eqs.~(\ref{e-genVs}) and
(\ref{e-spectVi}). The exponential factor on the r.h.s. gives the
probability for an electron to travel ballistically to a point of
scattering, after which it reaches the detector without further
collisions. In the local approximation of Eq.~(\ref{e-spectVi}),
$\Lambda = L \alpha / 4 \sin(\alpha / 2)$ can be interpreted as
the length of the trajectory from the injector to point $O$ (see
Fig.~2).

In order to compare the experimental data with theory, it is
necessary to extract the contribution of scattered electrons,
$\delta V_{d}^{s}$, from the observed signal $\delta V_{d}$. We
have
%%%%%%%%%%%%%%%%%% Eq. 7 %%%%%%%%%%%%%%%%%%%%%%%%%%%%%%
  \begin{equation}  \label{e-experVs}
    \delta V_{d}^{s}=\delta V_{d}- \exp\left(
     {-\frac{2r_c}{l_{ee}(V_i)}\arcsin \frac{L}{2r_c}}\right)\delta V_{d}^{0}\,.
  \end{equation}
%%%%%%%%%%%%%%%%%%%%%%%%%%%%%%%%%%%%%%%%%%%%%%%%%%%%%%%
Here, $\delta V_{d}^{0}(B)$ is the signal which would be observed
in the absence of scattering, so that the second term on the
r.h.s.\ of Eq.~(\ref{e-experVs}) is the contribution of electrons
that reach the detector ballistically. Experimentally, $\delta
V_{d}^{0}(B)$ can be obtained from the experiment at lowest
injection energy $V_{i}=0.1$ mV. In this case $l_{ee}/L \sim
10^{2}$ and thus collisions can be neglected.
%, at the same time $V_{i}\gg T$.
For $l_{ee}$ we use the expression for energy relaxation in a 2DEG
obtained by Giuliani and Quinn [cf.\ Ref.~\cite{guiliani82},
Eq.~(13)].

\begin{figure}[H]
\begin{center}
\resizebox{6.25cm}{10cm}{\includegraphics{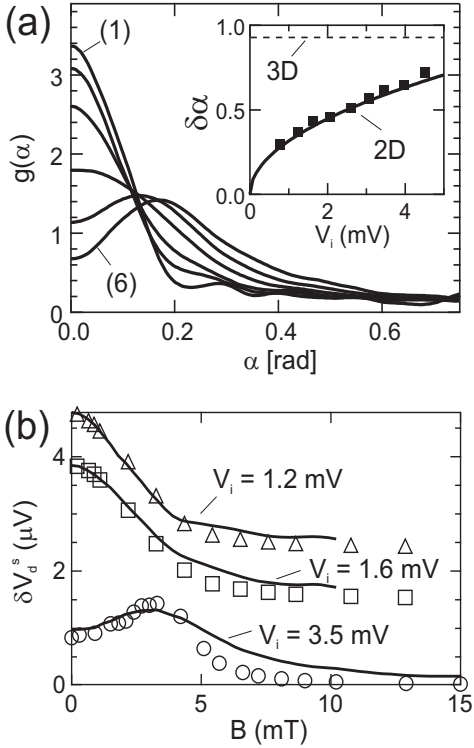}}
\end{center}
\caption{(a) ee-scattering distribution function $g(\alpha)$
restored from the experiment. Inset: width $\delta\alpha$ of
$g(\alpha)$ (squares) as function of injector bias voltage $V_i$.
$\delta\alpha$ is defined as the angle below which $2/3$ of the
electrons are scattered. Solid and dashed lines represent
theoretical prediction for a 2D ($\delta\alpha \propto
(eV_i/E_F)^{1/2}$) and a 3D system, respectively.
    (b) Comparison of the theoretical [markers, Eq.~(\ref{e-convVi})] and
experimental [lines, inferred from the experiment using
Eq.~(\ref{e-experVs})] $\delta V^s_d $, the detector signal due
only to scattered electrons. Curves are display with an offset for
clearness.}
    \label{g-verify}
\end{figure}

Fig.~\ref{g-verify} (a) shows the angular distribution functions
$g(\alpha)$ for various injector energies obtained from the
experimental data in Fig.~\ref{measurement} using
Eqs.~(\ref{e-spectVi}), (\ref{e-mocaG}), and (\ref{e-experVs}).
The various $g$ clearly display the expected small-angle
scattering behaviour. For small $V_i$ (curves (1) and (2), $V_i=$
0.8 and 1.2 mV, respectively) the observed peak width
$\delta\alpha$ is only slightly larger than the point contact
opening angle $\Phi$.  As discussed above, in this limit the
experimentally recovered $g(\alpha)$ is smoothed, and we cannot
expect to observe the dip at very small angles.

The peak in $g(\alpha)$ broadens when the energy of the injected
electrons is increased (curves 3-4). In the inset of
Fig.~\ref{g-verify} (a) the width  of $g(\alpha)$ is displayed as
a function of injection energy. It shows a clear square-root
behaviour $\delta\alpha\propto\sqrt{eV_i/\varepsilon_F}$ in
contrast to 3D where $g$ is essentially energy-independent. The
increase of $\delta\alpha$ with $V_i$ also directly implies that
the small angle scattering observed by us can not be attributed
with weak screening in 2D systems. In this case the scattering
angle should actually {\it decrease} with excess energy.

When $\delta\alpha$ becomes larger than the QPC opening angle
$\Phi$ for higher $V_i$, one can clearly observe the expected dip
in forward direction [curves 5-6, Fig.~\ref{g-verify} (a)]. The
amplitude of the dip is much larger than would be the case for a
3D electron system.

%%%%%%%%%%%%%%%%%%%%%% Fig. 4 %%%%%%%%%%%%%%%%%%%%%%%%%

As discussed above, the local approximation of
Eq.~(\ref{e-spectVi}) is not valid at small scattering angles
$\alpha<\Phi$. For these angles, $g$ is more precisely given by
the integral equation:
%%%%%%%%%%%%%%%%%% Eq. 9 %%%%%%%%%%%%%%%%%%%%%%%%%%%%%%
 \begin{eqnarray} \label{e-convVi}
   \delta V_{d}^{s}\simeq 2 C \nu \int d\varphi \rho_{i}(\varphi) \int
   d\varphi' \rho_{d}(\varphi') \times \nonumber\\
   \tilde{g}(\varphi'-\varphi + L /r_{c},
   V_{i}) \kappa \big(\varphi'-\varphi+L /r_{c}\big),
 \end{eqnarray}
%%%%%%%%%%%%%%%%%%%%%%%%%%%%%%%%%%%%%%%%%%%%%%%%%%%%%%%
where $\kappa(x) = 1/|x|$ for $x>\Phi$ and $\kappa(x) = 1/\Phi$
for $x<\Phi$. Here again we use $\tilde{g}(\alpha)$ as defined in
Eq.~(\ref{e-mocaG}); $\Lambda = L(2 \varphi' + L/r_c)/2(\varphi'-
\varphi+L/r_c)$ is the distance between injector and the crossing
point (O) of electron trajectories injected at angle $\varphi$ and
detected at angle $\varphi'$; the integration in (\ref{e-convVi})
has to be evaluated for all $\Lambda$ such that $0<\Lambda<L$,
while $l_{ee} \equiv l_{ee}(V_i)$.

For comparison the results of Eq.~(\ref{e-convVi}) for various
values of $V_i$ are presented together with experimental data for
$\delta V_{d}^{s}$ in Fig.~\ref{g-verify} (b). As is evident from
the figure, we find a gratiying agreement between theory (markers)
and experiment (drawn curves), justifying the assumptions made in
extracting $g$ from the experimental data.

The results presented in this paper demonstrate that a suitably
performed electron-beam experiment can provide a wealth of detail
on fundamental electron scattering processes, not only on the
energy dissipation, but certainly also on momentum scattering.
Note that by changing the detector gate voltage it is also be
possible to analyze the scattered electrons' energy dependence
\cite{Heiblum}. Combining this with $B$-field dependent data as
discussed in the present paper should enable a direct experimental
determination of the collision integral kernel $\nu_{\bf p p'}$ as
a function of $\varepsilon'$ (detection energy), $\varepsilon$
(injection energy, $eV_i$) and angle $\alpha$. Although in the
present paper we dealt solely with ee-collisions, one can
investigate in the same way other scattering processes, e.g.\
electron-phonon \cite{Heiblum} or electron-impurity collisions.

In conclusion, electron-beam experiments in the 2DEG of
GaAs-(Ga,Al)As heterostructures demonstrate unambiguously the
occurrence of small-angle ee-scattering. The scattering
distribution function broadens with increasing electron energy,
$\delta \alpha \propto \sqrt{eV_i / \varepsilon_{F}}$, and a
pronounced dip occurs at small angles. The observations represent
conclusive evidence for the manifestation of 2D density-of-states
effects in the ee-scattering process.

\acknowledgements

This work was supported in part by ``Volkswagen-Stiftung'' (Grant
No.I/72531) and by the DFG (MO 771/1-2).

%%%%%%%%%%%%%%%%%%%%%%%%%%
\end{multicols}
%%%%%%%%%%%%%%%%%%%%%%%%%%%

\end{document}